\documentclass{sig-alternate}
\newfont{\mycrnotice}{ptmr8t at 7pt}
\newfont{\myconfname}{ptmri8t at 7pt}
\let\crnotice\mycrnotice%
\let\confname\myconfname%

\usepackage[american]{babel}
\usepackage[T1]{fontenc}
\usepackage[utf8]{inputenc}
\usepackage{times}
\usepackage{textcomp}
\usepackage{graphicx}
\usepackage{amssymb}
\usepackage{microtype} 

\usepackage{url}
\renewcommand{\tt}{\fontencoding{OT1}\fontfamily{cmtt}\selectfont}

\usepackage[usenames,dvipsnames]{color}
\usepackage{booktabs}
\usepackage[numbers,sectionbib]{natbib}
\makeatletter
\def\@copyrightspace{\relax}
\makeatother

\begin{document}

\title{Approaching the Ad Placement Problem with Online Linear Classification}
\subtitle{The winning solution to the NIPS'17 Ad Placement Challenge}

\numberofauthors{1}
\author{
  \alignauthor
  Alexey Grigorev\\
  \affaddr{Simplaex GmbH}\\
  \affaddr{contact@alexeygrigorev.com}\\
  \alignauthor
}

\maketitle

\begin{abstract}

The task of computational advertising is to select the most suitable 
advertisement candidate from a set of possible options. The candidate 
is selected in such a way that the user is most likely to positively 
react to it: click and perform certain actions afterwards. 

Choosing the best option is done by a ``policy''~-- an algorithm which 
learns from historical data and then is used for future actions. This 
way the policy should deliver better targeted content with higher chances 
of interactions.

Constructing the policy is a difficult problem and many researches and
practitioners from both the industry and the academia are concerned with it.
To advance the collaboration in this area, the organizers of 
NIPS'17 Workshop on Causal Inference and Machine Learning 
challenged the community to develop the best policy algorithm. The challenge 
is based on the data generously provided by Criteo from the logs 
of their production system.

In this report we describe the winning approach to the challenge: our team 
was able to achieve the IPS of 55.6 and secured the first position. 
Our solution is made available on GitHub\footnote{\url{https://github.com/alexeygrigorev/nips-ad-placement-challenge}}.

\end{abstract}

\section{Introduction}

In computational advertisement the goal is to select the best possible 
advertisement for the user. Consider the following scenario: a user 
opens a website with advertisement slots (``impressions'').
Our system has many possible options (``a candidate set'') for filling these slots. 
To do it, a ``policy'' is leaned from the past interactions of users with 
the system. Such policy is then used to select the best candidate 
from the candidate set for the future visits to the website.

This problem is quite important for the online marketing community and many researches are concerned with it. Even a slight improvement
in the CTR (``Click-Through Rate'') will lead to revenue increase and 
better user experience.

As a part of NIPS'17 Workshop on Causal Inference and Machine Learning, the 
organizers prepared the Ad Placement task. In the task they challenged the 
Data Science community to develop a policy based on historical data. 
Criteo has generously donated
a large dataset with interactions from their production system
\cite{lefortier2016large}. Based on this data, the participants could 
approach this task and build policies to optimize the CTR.
The challenge was hosted at the \texttt{crowdai.org} 
platform\footnote{\url{https://www.crowdai.org/challenges/nips-17-workshop-criteo-ad-placement-challenge}}.
The solutions of the participants were evaluated using IPS~-- Inverse Propensity Score (see~\cite{lefortier2016large} for more details about evaluation). 

In this paper we present our approach to the challenge. 
Our team used the Follow-The-Regularized-Lead-Proximal (FTRL-Proximal) 
algorithm~\cite{ftrl} for solving the task. 
We show that for such large scale sparse classification task 
FTRL-Proximal is very competitive and was able to produce a high-scoring policy.
Our solution was able to achieve the IPS of 55.6, putting us at the first position of the competition leaderboard.

\section{Dataset Description}

The dataset provided for the competition has two parts: training and testing. 
The training file is used for learning the policy, which is then executed against 
the testing part and the results are sent back to the platform for evaluation.

The training dataset contains approximately 10.5 mln 
candidates set (3Gb compressed), with 12.3 candidates in each set on average. 
Additionally, we know which candidate was
selected by the system, the propensity score of the selected candidate 
and whether the user clicked at it or not. The CTR in the training data
is around 5\%. 

The test set contains 7 mln groups (1.4Gb compressed) and neither propensity information nor clicks is provided.

A typical candidate set looks like the following (the lines are truncated):

\begin{verbatim}
17193693 |l 0.999 |p 11.324800021|f 0:300 ...
17193693 |f 0:300 1:250 2:1 10:1 11:1 12:1 ...
17193693 |f 0:300 1:250 2:1 12:1 14:1 21:1 ... 
17193693 |f 0:300 1:250 2:1 12:1 14:1 21:1 ... 
17193693 |f 0:300 1:250 2:1 12:1 14:1 21:1 ... 
17193693 |f 0:300 1:250 2:1 12:1 14:1 21:1 ... 
17193693 |f 0:300 1:250 2:1 12:1 14:1 21:1 ... 
\end{verbatim}

Here for the group with \verb|id=17193693| the first candidate is 
selected by the system. Is has the propensity score 11.32 and was not clicked
(the label is \verb|0.999|). The clicked candidates have the label \verb|0.001|.
Each candidate is characterized by a set of features, stored 
in the \verb|feature:value| form.
More information about the dataset can be found in the dataset companion
paper by D.Lefortier et al \cite{lefortier2016large}.

In the next section we will describe our solution in more details.

\section{Approach}

In this section we present our approach to the challenge.
First, we describe the hardware and software used for the solution,
then we show the validation scheme used in our experiments,
and finally we talk about the features, the model we built on these features
as well as the post-processing scheme we used for modifying the model's output.

\subsection{Environment}

The experiments were performed on a Linux Ubuntu 14.04 server with 32GB RAM and 8 Cores.

We used Python 3.5.2 and the PyData stack for our development:

\begin{itemize}
  \item \texttt{numpy} 1.13.3 for numerical operations \cite{van2011numpy};
  \item \texttt{scipy} 0.19.1 for storing sparse data matrices \cite{jones2015scipy}.
\end{itemize}

We used Anaconda --~a distribution of Python with many
scientific libraries pre-installed\footnote{\url{https://www.continuum.io/downloads}},
including both the aforementioned libraries.

For online learning we used our own implementation of the FTRL-Proximal
algorithm\footnote{\url{https://github.com/alexeygrigorev/libftrl-python}}
which is available online and can freely be used by anyone.

\subsection{Modeling}

\textbf{Validation}

We split the training data into four parts by the id of the candidate set.
For training we used the parts 0, 1 and 2, and the 3rd part was used for validation.
The selected training part contained 10.6 mln candidate sets and remaining 3.5 mln
was left for the validation part. 

We did not hold out any extra data as a test set: for this purpose we used the 
provided test set and verified our score using the leaderboard of the competition. 

\ \\

\textbf{Data Preparation and Features}

The organizers have already extracted features from their production logs
and represented each candidate by a 74000-dimensional sparse vector. Most of the
features were already one-hot-encoded, but not all of them: many features have 
other values apart from 0's and 1's. 

To further process the data we tried the following approaches:

\begin{itemize}
\item disregard the value of each feature and always treat it as binary,
\item perform the hashing trick for feature + value to ensure it is one-hot-encoded \cite{weinberger2009feature}.
\end{itemize}

\ \\

\textbf{Online Learning Model}

Once the dataset was converted to a sparse matrix, we trained a supervised model. 
Each click was treated as a positive observation, no click~-- as a negative one. 

To perform the learning we chose the FTRL-Proximal algorithm~\cite{ftrl}: 
it has proved to be competitive in the computational advertisement settings
and has showed great performance for other sparse problems. For this challenge 
we used \texttt{libftrl-python}~-- our own implementation which is highly 
efficient in multicore environment and performs Hogwild!-style updates \cite{recht2011hogwild}, thus
resulting in very fast training time.

We had two options for data processing: with hashing and without. Our validation
did not show any significant difference between these two approaches and thus
we preferred the simpler one: treating everything as features with binary values.

However, we noticed that our model sometimes exhibits high variance, which 
then leads to a suboptimal policy. 
To overcome this issue we stabilized the model by 
training it multiple times and then taking the average prediction. 

\ \\ 

\textbf{Post-Processing}

Processing the output of the model was very important in our solution: 
it allowed to account for the behavior of the competition metric
which favored clicks predicted with high certainty. 

The IPS metric was computed in the following way:

\begin{itemize}
\item model scores are recorded,
\item the softmax function over all scores in the candidate set is computed,
\item the resulting score of the selected item is multiplied by its propensity,
\item only sets with a clicked item are considered, the ones with no clicks are ignored.
\end{itemize}

Thus is it very important to make sure that the best candidate receives most 
of the probability. This way it contributes higher score to the overall
metric.

To do it we used a two step post-processing procedure:

\begin{itemize}
\item apply the sigmoid function to each score, scale it by coefficient $C$,
\item add $M$ to the maximal value of the set.
\end{itemize}

The second step alone is not enough: when our model is not correct 
and selects a wrong candidate, this candidate gets most of the 
probability and does not give any contribution to the global score. 
However, the first step makes the re-arrangement of probabilities 
smoother and not drastic in cases of almost-tie candidates. 

The values of $C$ and $M$ were selected using our validation set.

\section{Evaluation Results}

Our final model is an average of 10 FTRL-Proximal models trained with the following parameters: 

\begin{itemize}
\item $\alpha=0.1$, $\beta=1$,
\item $L_1 = 75, L_2 = 25$,
\item $C = 850100, M = 15$.
\end{itemize}

This combination lead to the IPS of 55.6, which resulted in the first position
(see table~\ref{final-standings}: our team is in bold\footnote{\url{https://www.crowdai.org/challenges/nips-17-workshop-criteo-ad-placement-challenge/leaderboards}}).

\begin{table}
\centering
\begin{tabular}{|c|c|c|}
  \hline
\textbf{Team name}  &   \textbf{IPS} &   \textbf{std} \\
  \hline
\textbf{ololo}     & 55.6 & 4.1 \\
geffy     & 54.6 & 1.9 \\
Group     & 54.3 & 1.6 \\
atsky     & 54.1 & 3.0 \\
mortiarty & 48.6 & 1.2 \\
  \hline
\end{tabular}
  \caption{Top 5 participants of the challenge.}
  \label{final-standings}
\end{table}

\section{Conclusion}

In this paper we showed that FTRL-Proximal, an online linear 
classification algorithm, is still very competitive for 
large scale sparse problems like the Ad Placement Challenge and 
it outperformed many other approaches. 
We also showed that post-processing was an important step in
achieving good IPS.

{\raggedright
\bibliography{report}

\begin{thebibliography}{6}
\providecommand{\natexlab}[1]{#1}
\providecommand{\url}[1]{\texttt{#1}}
\expandafter\ifx\csname urlstyle\endcsname\relax
  \providecommand{\doi}[1]{doi: #1}\else
  \providecommand{\doi}{doi: \begingroup \urlstyle{rm}\Url}\fi

\bibitem[Jones et~al.(2015)Jones, Oliphant, Peterson, et~al.]{jones2015scipy}
E.~Jones, T.~Oliphant, P.~Peterson, et~al.
\newblock Scipy: Open source scientific tools for python, 2001.
\newblock \emph{URL http://www. scipy. org}, 73:\penalty0 86, 2015.

\bibitem[Lefortier et~al.(2016)Lefortier, Swaminathan, Gu, Joachims, and
  de~Rijke]{lefortier2016large}
D.~Lefortier, A.~Swaminathan, X.~Gu, T.~Joachims, and M.~de~Rijke.
\newblock Large-scale validation of counterfactual learning methods: A
  test-bed.
\newblock \emph{arXiv preprint arXiv:1612.00367}, 2016.

\bibitem[McMahan et~al.(2013)McMahan, Holt, Sculley, Young, Ebner, Grady, Nie,
  Phillips, Davydov, Golovin, Chikkerur, Liu, Wattenberg, Hrafnkelsson, Boulos,
  and Kubica]{ftrl}
H.~B. McMahan, G.~Holt, D.~Sculley, M.~Young, D.~Ebner, J.~Grady, L.~Nie,
  T.~Phillips, E.~Davydov, D.~Golovin, S.~Chikkerur, D.~Liu, M.~Wattenberg,
  A.~M. Hrafnkelsson, T.~Boulos, and J.~Kubica.
\newblock Ad click prediction: a view from the trenches.
\newblock In \emph{Proceedings of the 19th ACM SIGKDD International Conference
  on Knowledge Discovery and Data Mining (KDD)}, 2013.

\bibitem[Recht et~al.(2011)Recht, Re, Wright, and Niu]{recht2011hogwild}
B.~Recht, C.~Re, S.~Wright, and F.~Niu.
\newblock Hogwild: A lock-free approach to parallelizing stochastic gradient
  descent.
\newblock In \emph{Advances in neural information processing systems}, pages
  693--701, 2011.

\bibitem[Van Der~Walt et~al.(2011)Van Der~Walt, Colbert, and
  Varoquaux]{van2011numpy}
S.~Van Der~Walt, S.~C. Colbert, and G.~Varoquaux.
\newblock The numpy array: a structure for efficient numerical computation.
\newblock \emph{Computing in Science \& Engineering}, 13\penalty0 (2):\penalty0
  22--30, 2011.

\bibitem[Weinberger et~al.(2009)Weinberger, Dasgupta, Langford, Smola, and
  Attenberg]{weinberger2009feature}
K.~Weinberger, A.~Dasgupta, J.~Langford, A.~Smola, and J.~Attenberg.
\newblock Feature hashing for large scale multitask learning.
\newblock In \emph{Proceedings of the 26th Annual International Conference on
  Machine Learning}, pages 1113--1120. ACM, 2009.

\end{thebibliography}
}
\end{document}